# Measuring Cellular Ion Transport by Magnetoencephalography

Sudhir Kumar Sharma,[a] Sauparnika Vijay,[b] Sangram Gore,[b] Timothy M. Dore,[b,c]* and Ramesh Jagannathan[a]*

[a] Engineering Division, New York University Abu Dhabi, UAE

[b] Science Division, New York University Abu Dhabi, UAE

[c] Department of Chemistry, University of Georgia, Athens, Georgia 30602, USA

**ORCID ID**

Sudhir Kumar Sharma: 0000-0003-2437-5934

Sangram Gore: 0000-0003-4236-7604

Timothy M. Dore: 0000-0002-3876-5012

Ramesh Jagannathan: 0000-0003-0269-6446

**Abstract**

The cellular-level process of ion transport is known to generate a magnetic field. A non-invasive magnetoencephalography (MEG) technique was used to measure the magnetic field emanating from HeLa, HEK293 and H9c2(2-1) rat cardiac cells. The addition of a non-lethal dose of ionomycin to HeLa and capsaicin to TRPV1-expressing HEK293 cells, respectively, resulted in a sudden change in the magnetic field signal consistent with $Ca^{2+}$ influx, which was also observed by confocal fluorescence microscopy under the same conditions. In contrast, addition of capsaicin to TRPV1-expressing HEK293 cells containing an optimum amount of $Ca^{2+}$ channel blockers, a TRPV1 antagonist (ruthenium red), resulted in *no* detectable magnetic or fluorescent signals. These signals confirmed that the measured MEG signals are due to cellular ion transport through the cell membrane. In general, there is evidence that ion channel/transporter activation and ionic flux are linked to cancer; Therefore, our work suggests that MEG could represent a non-invasive method for detecting cancer.

**Introduction**

At a global level, cancer therapeutics suffer the lowest success rate compared to therapeutics for other major diseases. All indications suggest that cancer will soon be the leading cause of mortality in developed countries. The future of cancer research depends on the successful inter-dependencies between therapeutic, diagnostic, and prognostic technologies [1,2]. The power of early detection in particular is key to successful therapeutic and prognostic outcomes, and non-invasive imaging techniques such as CT, MRI, and PET are proving to be extremely useful [1,3]. A radically different approach to early-stage detection would focus on the development of non-invasive techniques to monitor the cellular-level process of ion transport through membranes and related membrane protein interactions as an indicator of cell health. The changes to membrane polarizations are measured by electrophysiology, fluorescent dyes and proteins that respond to changes in voltage or specific charged analytes (e.g., $Ca^{2+}$) in the cytoplasm. There is evidence that ion channel and transporter activation are linked to cancer [4–6]. Dividing cells like HeLa cells have a lower membrane potential (-48 mV) [7] than non-dividing ones (e.g., neurons -60 to -70 mV) [8]. Metastatic cancer cells are typically more depolarized than normal cells [9,10]. Overexpression of potassium channels have been strongly associated with a number of cancer cell lines, and the phenomenon is generally accepted as fundamental to understanding cancer biology [4,11,12].

From basic physics, it is known that an ion (i.e., a charge carrier) moving through an electric field will generate a current and a magnetic field. Naturally, it follows that ion flow through polarized cell membranes should result in a net magnetic field and magnetic sensors, in principle, have the potential to quantitatively measure them. Detectors of magnetic fields have been used to probe nerve impulses and ion transporters in cell and tissue culture [13–15]. Wijesinghe described the results of work with a specially designed neuro-magnetic current probe to measure magnetic fields created by single axons and bundles [16]. Precise, high-resolution measurements of action potentials from individual neuronal activity have been made using a technique based on optically probed nitrogen vacancy (NV) quantum defects in diamond [17,18]. This technique enabled long-term data collection without any bleaching effects at sub-millisecond resolution, and is label free. However, it did not have sufficient detection sensitivity for weak magnetic fields that are generated through cell membranes in broad classes of different cell types.

Magnetoencephalography (MEG) was one of the three early techniques that clearly established the existence of measurable magnetic fields due to ionic action currents in biological tissues [19–21]. Due to significant advances in computational science, MEG has become a useful tool to study brain function, neuronal activities, and their associated magnetic fields. In the last several years, significant research activity in the MEG field has been able to firmly establish the quantitative, causal relationship between magnetic signals and ionic currents in isolated nerve axons and muscle systems [13]. A MEG instrument consists of an array of super conducting quantum interference device (SQUID) detectors [22,23]. They can measure fields as small as a femtotesla in millisecond time frames. Modern-day MEG systems consist of axial gradiometers that only measure a change in magnetic field.

Taken together, these observations suggest that SQUID devices arrayed in a MEG instrument, may have the potential to reveal fundamental information about cancer by non-invasively measuring the magnetic signature of the tissue. [22,23]. In this manuscript, we propose that the naturally emanating magnetic signals generated by ion flux in various cell types in culture can be quantitatively measured using the MEG system.

**Results and Discussion**

We adapted a MEG system originally designed for a human head and instead applied it to tissue culture. Data were collected on a 224-channel (sensors) MEG system (Model: PQ1128R, Figure 1a) with a headpiece designed to fit a human subject's head (Figure 1b) containing an array of 208 SQUID sensors (Figure 1c), and 3 orthogonally oriented reference sensors located away from the headpiece. It is important to note that, at a given time, the MEG system consisting of axial gradiometers will *only* measure a change in a contiguous magnetic field that is perpendicular to the pickup coil. In otherwords, the MEG signals describe the temporal characteristic of ion transport through the channels due to opening and closing of the channels (i.e. the rise and fall rate) and could provide valuable information regarding channel gating dynamics. A zero MEG signal might imply a constant, steady-state trans-membrane ionic flux. Various parameters were examined, such as cell type (HeLa, HEK 293 and H9c2 (2-1)), number of cells ($n = 1 \times 10^5$, $5 \times 10^5$, $1 \times 10^6$, and $2 \times 10^6$), flask sizes, shapes, and location and orientation within the MEG headpiece, to determine the best experimental protocol for our experiments. We decided to use T-25 flasks as a preferred choice and maintained all the geometrical parameters relative to the MEG headpiece constant. We also took a more constrained approach to only include signals equal to 750 fT or greater in magnitude.

*Ionomycin induced Ca$^{2+}$ flux in HeLa cells in culture.*

*1. Confocal Fluorescence Microscopy Studies.*

Ionomycin is a diacid polyether antibiotic that has a high affinity for Ca$^{2+}$, which gives it ionophoric properties [24]. Ionomycin stimulates release of Ca$^{2+}$ from internal stores and Ca$^{2+}$ influx from the extracellular space through ion channels, or both[25,26]. The net effect is an increase in intracellular [Ca$^{2+}$]. The smallest dose of ionomycin to give a Ca$^{2+}$ signal from the Ca$^{2+}$-sensitive fluorescence dye (Fluo-4) was determined by treating HeLa cells in culture with increasing ionomycin concentrations and observing the effects by confocal fluorescence microscopy (Figure 2a). Below 2 µM, ionomycin did not result in any noticeable fluorescence signal, but maximal signal was observed at 2 μM, indicating $Ca^{2+}$ influx into the cytosol. Exposure to high ionomycin concentrations caused extended cell death. We determined the optimum concentration of ionomycin and length of exposure to be 1, 2, 5 μM and 5, 10, 30 min, respectively. The extent of cell death was measured by trypan blue assay (Figure 2b). A 2-μM ionomycin concentration for an exposure of 5 minutes resulted in 10% of cell death, and longer exposure times and higher concentrations led to greater cell death (Figures 2c and 2d).

*2. MEG Studies*

We repeated the same fluorescence microscopy experiment using a MEG instrument to measure the change in magnetic field. The addition of 2 μM ionomycin to $1 \times 10^6$ HeLa cells resulted in a change in the magnetic signal (Figures 2e and S1a). It took approximately 1-2 seconds to add all the ionomycin solution to the culture flask and record the time of completion. The red dotted line represents the end of the ionomycin solution addition to the culture flask. Immediately afterwords, a sudden, significant increase in the magnetic signal ( 50-75 pico tesla) in several channels was observed over a period of 0.15 s and signal rapidly decayed thereafter. It is reasonable to assume that the processes related to the release of $Ca^{2+}$ from internal stores and $Ca^{2+}$ influx from the extracellular space through ion channels, were triggered from the moment we started to add ionomycin to the culture flask.  Since the MEG system only detects a change in the magnetic field and inomycin induced $Ca^{2+}$ flux is irreversible, the rapid decay of the MEG signal after 0.15 second probably indicated the onset of a steady-state $Ca^{2+}$ flux.  During the signal oscillations, the intensities of the signals were attenuated, probably due to saturation of the sensors. Over-lay plots of the signals measured by channels 121,127, and 142 (Figure S1b) show that the temporal response rates and curve shapes of the signals are identical; this confirms that the source of the signals is the same. The signal measured by the channel 121, does not show any spikes, probably because the orientation of the spiked response is not perpendicular to the pickup coil of the sensor or perhaps because the sensor is located further away from the cells. For the control experiment, we repeated the above experiment but added the same volume of culture medium without any ionomycin to the flask. We found no detectable change in the magnetic signal (Figure S1c).

*Capsaicin induced $Ca^{2+}$ flux in TRPV1-expressing HEK293 cells in culture.*

*1. Confocal Fluorescence Microscopy Studies.*

*2. TRPV1-expression in HEK293 cells:*

A method to induce $Ca^{2+}$ ion flux is to activate transient receptor potential cation channel sub-family V member 1 (TRPV1) on the surface of cells. The TRP family of ion channels are well-understood cellular sensors that regulate response to temperature, touch, pain, and other stimuli[27,28]. Activation of TRPV1 by either binding of a ligand such as capsaicin, a small molecule that is the active component in chili peppers and imparts a burning sensation by activating nociceptive sensory neurons; *N*-vanillyl-nonanoylamide (VNA), an equipotent capsaicin analog; or by exposure to noxious heat (>37 °C) results in nerve terminal depolarization and generation of action potentials[29]. The responses observed by engineered and endogenously expressed TRPV1 channels to both applied capsaicin and exposure to heat are nearly identical[30], making activation of TRPV1 channels a versatile method for studying signal transduction activity of sensory neurons.

Using a confocal fluorescence microscope, we measured the influx of $Ca^{2+}$ with the $Ca^{2+}$ sensitive dye (Fluo-4) before and after the TRPV1-expressing HEK293 cells were exposed to the TRPV1 receptor agonist capsaicin (Figure 3a). Capsaicin dosage levels were optimized to achieve maximum fluorescence signal with minimal toxicity (Figure S2a1). In these experiments, activation of TRPV1 channels depends on the diffusion of capsaicin to the cell receptors. Capsaicin (10 μM final concentration) was added to these cells and the influx of $Ca^{2+}$ was observed (Figures 3b and S2a2). The cells responded with a significant increase in fluorescence signal after the addition of capsaicin. The fluorescence signal in some cells decayed back to nearly starting levels, whereas in other cells this signal decay was not observed. This is probably because there is ample capsaicin in the media to keep the ligand bound to the channels which hold them open. The cell surface expression of TRPV1 was verified by Western blot and immunohistochemistry (Figure S2b). In another series of experiments, we investigated the effect of the TRPV1 antagonist ruthenium red on capsaicin addition to TRPV1-expressing HEK293 cells. We added ruthenium red (10 μM final concentration) to the flask containing adherent HEK293 cells prior to the addition of capsaicin (Figure 3c). As expected, subsequent addition of capsaicin (10 μM final concentration) did not result in any detectable fluorescence signals (Figure 3d and S2a3).

*3. MEG Studies*

The MEG experiments with TRPV1-expressing HEK293 cells were carried out in a similar manner to the fluorescence microscopy experiments. We added capsaicin (10 μM final concentration) to HEK293 cells in culture and observed a significant spike in magnetic signal in several channels (Figure 3e). The black dotted lines in Fig.3e represent the approximate time of completion of the capsaicin solution addition to the culture flask. The signals decayed to the baseline in a way similar to the experiments with ionomycin addition. The intensity of the spike in magnetic signal due to capsaicin ( 3-5 pico tesla) was much less than that obnserved in the ionomycin experiments ( 50-75 pico tesla) and the much shorter time duration of the spike was probably due to the magnetic signal strength falling below our threshold of 0.75 pico tesla. Over-lay plots of these signals (Figure S2c) confirmed that the signals measured by these channels originated from a single source. Plotting the dependence of the measured magnetic field signal as a function of [capsaicin] (Figure S2d) revealed that increasing [capsaicin] from 0 to 10 μM resulted in a proportional increase in [$Ca^{2+}$] which reached a steady state value. Overall, it is clear that an increase in [capsaicin] and hence an increase in [$Ca^{2+}$] resulted in an increase in the magnetic field signal. The effect of the addition of capsaicin to adherent, TRPV1-expressing HEK293 cells detected by an immediate spike in the magnetic signal, has been repeated several times for different concentrations. These results strongly support our hypothesis that the measured magnetic signals correspond to cellular ionic flux. In control experiments, when only culture medium was added to the cells, we detectable no increase in magnetic signal (Figure S2e). We also investigated the effect of TRPV1 agonist (ruthenium red) on capsaicin addition to TRPV1-expressing HEK293 cells in culture. Ruthenium red (10 μM final concentration) was added to the flask containing the adherent HEK293 cells in culture prior to the addition of capsaicin. No detectable magnetic signals were observed after the capsaicin (10 μM final concentration) addition (Figure 3f). The experiment was repeated several times during the same day and on different days. The results are consistent with the fluorescence experiments, demonstrating that the magnetic signals measured by the MEG system are strongly correlated to the cellular ionic flux. We would like to note that while we are monitoring the [$Ca^{2+}$] activity, a major controibutor to the ionic flux through these channels would be [$Na^+$].

*Effect of multiple additions of capsaicin.*

    *1. <u>Confocal Fluorescence Microscopy Studies</u>*

We further examined the stability of the adsorbed capsaicin on the cell membrane by making multiple additions of capsaicin to the same cells and observing the effect with confocal microscopy (Figure S2f). Whereas the first addition of capsaicin (10 μM final concentration) resulted in a spike in the fluorescence intensity, the second addition did not result in any detectable response. This observation likely results from saturation of the TRPV1 receptors with capsaicin, which has an EC$_{50}$ of approx. 300 nM [31].

2. *MEG Studies*

The MEG experimental results were also found to be consistent with fluorescence experiments when capsaicin was added at 100 s and 725 s (Figure S2g) to the same cell population. That is, whereas the first addition of capsaicin (10 μM final concentration) resulted in a spike in the magnetic signal, the second addition of the same concentration did not result in any detectable response. Similar observations were made when the experiments were repeated with different waiting periods of up to 10 minutes between the two capsaicin additions.

In a second set of experiments, we examined the effect of replacing the culture before the second addition by washing the cells with fresh culture three times (Figure S2h). Once again, the first addition of capsaicin resulted in a magnetic signal, but no detectable magnetic response was observed after the second addition (Figure S2i), probably because the TRPV1 channels were saturated with the agonist. Any subsequent addition of capsaicin would therefore have no significant effect on Ca$^{2+}$ flux through the channels. The results support the hypothesis that the measured magnetic signals are due to the cellular ionic flux.

*Magnetic field measurements from cells in culture.*

*HeLa cells.*

MEG data from channels 3, 21, 35, 45, and 175 for $1 \times 10^6$ HeLa cells in culture at 80% confluency reveal a cluster of signals, approximately 1-3 pico tesla (pT) in intensity (Figure 4a and S3a). Overlay plots of signals from three channels (3, 21, and 45) for a period of 0.1 s (Figure S3b) indicate that the temporal response rates and curve shapes of these signals are identical with respect to each other, implying that all the channel signals originate from the same source. We therefore hypothesize that these weak signals are due to the normal trans-membrane ionic flux and that each signal wavelength (FWHM) correspond to the rise and fall rate of ion flux during a channel opening and closing. Fast Fourier Transform (FFT) of the data from channel 175 revealed a high level of periodicity in these signals with a characteristic frequency of 27.8 Hz ($36 \times 10^{-3}$ s) (Figure S3c). If assume that each adjacent signal correspond to an open and closed state, our data would imply that, at any given time, HeLa ion channels either stay opened and closed for a duration of $36 \times 10^{-3}$ s, respectively.

Two series of control experiments were carried out to establish the system noise floor. In the first experiment, we measured the magnetic field from a flask containing only culture, but no cells (Figures 4b and S3d). No detectable signals were observed. In the second experiment, we measured magnetic fields from HeLa cells dispersed in culture (Figure S3e). Once again, we did not measure any detectable magnetic signals. This is consistent with our expectation that in the case of cell dispersions in the culture there would be no net magnetic field due to symmetry considerations. This control experiment also eliminates any concern of signal contribution/contamination from any unknown "cell related debris" in the culture.

### _Non-differentiated H9c2(2-1) rat cardiac myoblasts vs differentiated myocytes._

The morphology of H9c2(2-1) cells were characterized before and after differentiation[32] (Figure S4). MEG signals from H9c2(2-1) rat cardiac myoblasts and differentiated myocytes were measured in a manner similar to that of the HeLa cells (Figure S5). Magnetic signals emanating from non-differentiated cells (Figure S5a) were similar to those of the HeLa cells, whereas differentiated cells exhibited clusters of signals (Figure S5b-c). FFT of the data, showed a high degree of periodicity for both cell lines (Figure S5d-f). Non-differentiated cell lines had a characteristic magnetic signature of 27.8 Hz, similar to the HeLa cells. In general, the differentiated cell lines consistently showed a cluster of higher frequency signals with a characteristic frequency around 220 Hz ($4.54 \times 10^{-3}$ s). If we assume that our earlier hypothesis for the HeLa cells is applicabable to other cell types, then we would conclude that the ion channels in the non-differentiated cell lines exhibited similar behavior and stay opened or closed for the same duration of time of $36 \times 10^{-3}$ s. The differentiated cell lines, however, stayed open or closed for a much shorter duration of time, namely, $4.54 \times 10^{-3}$ s. The delineation of different frequencies within the cluster of signals in the differentiated cell lines is likely due to different specific ion channels.

**Conclusion**

We report a non-invasive method to measure the net ionic transport through an electric field at a polarized membrane using MEG system. Addition of a non-lethal dose of ionomycin to HeLa or capsaicin to TRPV1-expressing HEK293 cells resulted in a sudden change in the magnetic signal which was consistent with confocal fluorescence experiments. Our experimental results with the TRPV1 agonist ruthenium red, which blocks the TRPV1 ion channels, showed that the magnetic signals detected by the MEG system are strongly correlated with the cellular ionic flux. It is interesting to observe that the non-differentiated H9c2(2-1) rat cardiac myoblasts and the differentiated myocytes showed a significant difference in their magnetic signatures. The characteristic frequency of 27.8 Hz for the myoblasts was found to be identical to that of HeLa cells, which is a cancer cell line. The differentiated cell lines showed a cluster of higher characteristic frequencies around 220 Hz. It is speculated that each of the three frequencies in the cluster is attributable to a specific ionic flux. One could then infer that the ion channel gating dynamics of non-differentited cell lines are non-distinct, a key factor affecting cellular homeostasis. Detailed investigations are still needed to correlate the magnetic signals to flux of specific ions, such as $Na^+$, $K^+$, and $Ca^{2+}$. This would further enable correlations to specific biological processes related to cell state, type, and differentiation stage. The resulting insights into the cellular processes at the membranes would potentially lead to a non-invasive method for early stage cancer detection.

**Experimental Section**

*Cell culture.*

HeLa cells were cultured in RPMI 1640 media with 10% FBS and 1% penicillin/streptomycin in cell culture flasks at 37 °C and 5% $CO_2$. HEK293 cells stably expressing TRPV1 were created by transfection with a pCMV6-NEO-TRPV1 plasmid (Origene) using TrueFect reagent (United BioSystems) and G418 (600 μg/mL) for selection. The cells were cultured in in DMEM media with 10% FBS and 1% PEN-STREP. G418 antibiotic (300 μg/mL) was used for maintaining the stable expression of TRPV1. TRPV1 expression was verified by Western blot and immunofluorescence staining using a TRPV1 antibody[33] (Figure S2b). H9c2(2-1) (ATCC® CRL1446™) cells were cultured in RPMI 1640 media with 10% FBS and 1% penicillin/streptomycin in cell culture flasks at 37 °C and 5% $CO_2$ at no greater than 50% confluence. To initiate differentiation, cells were allowed to grow to confluency and maintained in 2% FBS supplemented media for four days [34].

*Calcium dye loading.*

Calcium dye (50 µg, Fluo-4 AM, Life Technologies) was dissolved in DMSO (50 mL). Both Fluo-4 AM and Pluronic F-127 (Molecular Probes) were added to HBSS resulting in a 0.002% final concentration of each. The mixture was sonicated for 5 min, loaded onto cells growing in a 35-mm glass bottom dish, and then de-esterified for 30 min in a humidified $CO_2$ incubator (37 °C, 5% $CO_2$).

*Confocal microscopy.*

Live cell imaging was carried out on an Olympus FluoView FV1000MPE confocal microscope. Excitation using an argon ion laser was set at 488 nm and emitted light was reflected through a 500-600 nm filter from a dichroic mirror. Data capture and extraction was carried out with FluoView 10-ASW version 4.0 (Olympus), Image J-Fiji[35], and DeltaGraph (Red Rock Software). Stock solutions of ionomycin and capsaicin were prepared in DMSO, and DMSO accounted for no more than 0.2% of the final concentration of these reagents. To initiate the experiments, ionomycin (2 µM final concentration) or capsaicin (10 µM final concentration) was added to the culture dish from a pipette.

*Evaluation of ionomycin-dependent cell death.*

HeLa cells grown in culture were treated with ionomycin solutions (1, 2, and 5 µM final concentrations) for a period of 5, 10, or 30 min. After the designated time period, the cell suspension was treated with trypan blue (final concentration of 0.4%) and immediately loaded into a hemocytometer. Dead cells are stained blue and live cells are unstained. The cells were counted manually using a handheld tally counter.

*Evaluation of capsaicin-dependent cell death.*

HEK293 cells stably expressing TRPV1 grown in culture were treated with DMSO (0.2%) or capsaicin solutions (1.25, 2.5, 5, and 10 µM final concentrations) for a period of 30 min. After the designated time period, the cell suspension was treated with with CellTiter Blue (Promega) and the fluorescence was measured after 1.5 h using 560/590 nm filters on a BioTek Synergy microplate reader and compared to an untreated control.

*MEG data acquisition and analysis.*

T-25 culture flask containing plated Hela/ HEK293/ H9c2(2-1) cells was placed inside the MEG chamber in a pre-determined position on the base of the helmet. MEG data was collected in a Continuous mode with a 208-channel axial gradiometer system (Kanazawa Institute of Technology, Kanazawa, Japan) using MEG 160 software for data collection and post processing. This MEG software is produced jointly by Yokogawa Electric Corp., Eagle Technology Corp., and the Kanazawa Institute of Technology (KIT). A low pass of 200 Hz and a high pass of 0.1 Hz and a sampling rate of 3000 Hz for data recording were used. Data from MEG system was acquired by a DAQ system assembled by Eagle Technology. For MEG data post processing, offline noise reduction with a specifically designed algorithm, i.e., the continuously Adjusted Least- Squares Method (CALM), using three reference magnetometers mounted outside the MEG head, was used[36]. The CALM noise reduction method eliminates any detected covariance between the measured data in the MEG sensors with respect to the reference sensors. This technique was specifically designed to eliminate low frequency (< 10 Hz) noise as well as large extramural unintentional magnetic noise to disrupt MEG sensors. The selected data plotting and analysis was done using Origin Pro 2016 software.

*MEG experiments for remote addition of ionomycin and capsaicin.*

The experimental set up was designed such that we could add the optimum dosage of ionomycin or capsaicin dissolved in culture to the culture flask remotely. We used a fine plastic tubing, one end of it being attached to the end cap of the culture flask. The other end of the tubing located outside the MEG room was attached to a plastic bulb which contained the ionomycin or the capsaicin solution. This arrangement enabled undisturbed measurement of the baseline signal from the cells in the flask and after the addition of ionomycin or capsaicin by applying pressure to the liquid inside the tubing through a plastic bulb attached to the other end, located outside the MEG room. The complete addition of ionomycin or capsaicin took about 1-2 sec".

<u>*Noise floor Characterization*</u>:

1. During the first phase of our experiments (2016), we carried out a detailed study of the noise floor of the MEG system and the contiguous environment. We worked closely with other active NYUAD faculty from the neurobiology department who relied on the MEG system for their core research, and the representatives of the company that built the MEG system. The purpose of these conversations was to get educated and trained on the capabilities/limits of the MEG system and adapt the experimental protocols associated with "good" data acquisition practices, identify potential sources of noise that are specific to the MEG system and their characteristic signatures. Overall, the MEG system was set up such that the noise floor is comparable to MEG systems in other global research facilities.

2. Our standard experimental procedure involved the following: all (reference) control experiments for specific set of studies and general background noise measurements were repeated during each experimental session. We measured the *noise floor* before and after each experiment and many times, measured the noise floor between experiments as well. For (reference) control experiments, we did a minimum of two repeats and sometimes up to 4 repeats in each session.

3. As described in the manuscript, we only considered signals above a significant threshold that is repeatable within each experimental session and are repeatable in other experiments carried out on different days.

4. For each experimental set, we usually measured signals from 3-4 cell cultures per session and collected 2-5 sets of data for each cell culture.

*Reference Controls:*

1. <u>*Neat Cell cultures*</u><u>:</u> As described in the manuscript, the control experiments for the "neat" *adherent* cell culture measurements involved the cell culture flask prepared in the same manner as with the *adherent* cells but *without* the cells. The data acquisition procedure was identical to that with the cells.

2<u>. *Ionomycin experiments:*</u> For the ionomycin experiment, the control experiment involved a procedure that was identical to that of the main experiment in which ionomycin in culture media (2 μM final concentration) was added to the cells using plastic tubing and pneumatic pressure, except for the following: *no* ionomycin was present in the culture for the control experiment. This was repeated during the experimental session on the same day and all the experiments were repeated again several times on different days.

3. *Capsaicin experiments:* For the capsaicin experiment, the control experiment involved a procedure that was identical to that of the main experiment in which 10 μM of capsaicin in culture medium was added to the adherent HEK cells using a plastic tubing and pneumatic pressure, except for the following: *no* capsaicin was present in the culture medium for the control experiment. This was repeated during the experimental session on the same day and all the experiments were repeated again several times on different days.

*4. Signal Processing and Statistical Analysis.* We did extensive statistical analysis of the data for non-differentiated cells. The time domain magnetic signals from select channels were transformed into the frequency domain using the FFT algorithm in MatLab and Origin Pro 2016. FFT results obtained from 9 individual sensors per experiment were selected for statistical testing. We calculated the standard error of the mean (S.E.M) for an average of 5-independednt experiments using Prism 6.0 software (GraphPad Software, Inc. La Jolla, CA, USA). Statistical significance between or among experiments was assessed by one-way analysis of variance (ANOVA) followed by Tukey's or Dunnett's post hoc test. A value of $p < 0.0001$ was considered to be statistically significant in our experiments.

*5. Data Availability*

All data generated or analyzed during this study are included in this published article (and its supplementary information files).

**Acknowledgments**

We acknowledge the valuable support provided by the MEG Lab group and the Core Technology Platforms at NYU Abu Dhabi for the use of the instruments and discussions. We thank Prof. Kartik Sreenivasan at NYU Abu Dhabi and Dr. Daisuke Oyama at Kanazawa Institute of Technology for their valuable insights in the analysis of the MEG data; Dr. Amitabha Mazumder (previously at NYU Langone Medical School) and Profs. George Shubeita and Dipesh Chaudhury at NYU Abu Dhabi for their support and discussions; Dr. Xi Wei for help with the early experiments; and Mr. Graham Flick for discussions about MEG data analysis.

**Supporting Information**

**Figure S1a:** Screen shot of noise reduced data shown in Figure 2e.

**Figure S1b:** Overlay plots of data shown in Figure 2e.

**Figure S1c:** Control experiment for the data shown in Fig. 2e. Culture medium added remotely to the cell flask.

**Figure S2a:** (1) Effect of addition of capsaicin (10 μM final concentration) to HEK293 cells expressing TRPV1 channels. (2) Effect of addition of capsaicin (10 μM final concentration) to HEK293 cells expressing TRPV1 channels pre-added TRPV1 agonist (10 μM ruthenium red).

**Figure S2b:** Verification of TRPV1 (MW = 93-95 kDa) expression in stably transfected HEK 293 cells. (a) Western blot analysis of wild type (WT) and stably transfected (ST) cells and a molecular weight ladder (L): protein extracts were loaded on 10% SDS polyacrylamide gels, then the nitrocellulose membranes were split at the position of proteins of 40 kDa. The antibodies used and their dilutions were anti-TRPV1 (1:500, rabbit polyclonal, Origene, cat. #TA309921), anti-biotin (1:1,000, HRP linked), and anti-rabbit-HRP (1:2,000, goat secondary). β-Actin protein (MW = 42 kDa) was used as a control. (b and c) Immunofluorescence staining of HEK 293(TRPV1) cells treated (b) with TRPV1 specific antibody (anti-TRPV1, 1:250; secondary antibody, 1:500) and (c) without primary antibody.

**Figure S2c:** Overlay plots of data shown in Fig, 3e.

**Figure S2d:** Effect of increasing [capsaicin] concentration on the measured magnetic field from TRPV1-expressing HEK293 cells.

**Figure S2e:** Control experiment for the data shown in Fig. 3e. Effect of addition of culture without capsaicin remotely to the cell flask.

**Figure S2f:** Confocal fluorescence experimental data on the effect of multiple additions of capsaicin to the same TRPV1-expressing HEK293 cells: (a) before capsaicin addition, (b) after capsaicin addition, (c) after second capsaicin addition, (d) fluorescence (normalized) as a function of time, after the first capsaicin addition and after the second capsaicin addition.

**Figure S2g:** MEG noise reduced data on the effect of multiple additions of Capsaicin to the same HEK293 cells expressing TRPV1channels.

**Figure S2h:** Confocal fluorescence experimental data on the effect of replacing the culture before the second capsaicin addition, by washing the with fresh culture three times: (a) before first capsaicin addition, (b) after first capsaicin addition, (c) before second capsaicin addition, (d) after

second capsaicin addition, (e) fluorescence intensity vs time after first addition (screen shot), (f) fluorescence vs time after second addition (screen shot).

**Figure S2i:** MEG data on the effect of replacing the culture before the second capsaicin addition, by washing the TRPV1-expressing HEK293 cells with fresh culture three times. a) Screen shot of the noise reduced data for first capsaicin addition.  b) Screen shot of noise reduced data for the second capsaicin addition.

**Figure S3a:** Screen shot of the MEG noise reduced data shown in Fig. 4a.

**Figure S3b:** Overlay plots of data shown in Fig. 4a.

**Figure S3c:** FFT of the data of channel 175 shown in Figure 4.

**Figure S3d:** Screen shot of the MEG noise reduced signals from the control experiment data shown in Fig.4b.

**Figure S3e:** Magnetic fields from HeLa cells dispersed in culture

**Figure S4: Morphological characterization of H9c2(2-1) cells before and after differentiation.** (a) Representative images of non-differentiated H9c2(2-1) cells grown in 10% FBS media showing the characteristic stellar shape of myoblasts. (b) Representative images of H9c2(2-1) cells after five days of differentiation in low serum media (2% FBS) showing the characteristic elongated shape of myocytes.

**Figure S5: Comparison of the magnetic signals from non-differentiated and differentiated cardiac cells.** (a) Data for $1 \times 10^6$ H9c2(2-1) rat cardiac myoblasts detected by channels 4, 21, 35, 166, and 190. (b and c) Data for $1 \times 10^6$ H9c2(2-1) rat cardiac cells differentiated into myocytes detected by channels 31, 35, 82, 97, and 166 and 4, 31, 93, 99, and 206. (d) FFT of the data for non-differentiated cells (channel 4). (e and f) FFT of the data for differentiated cells (channels 97 and 206).

**Competing Interests**

The authors declare no competing interests.

**Contributions**

SKS and SV contributed equally to this work. RJ developed the concept of measurable magnetic fields emanating from cellular ionic transport using MEG and the potential distinction in magnetic signatures between normal and cancer cells. TMD designed and interpreted the microscopy experiments. RJ and SKS designed the MEG experiments and SKS conducted the MEG experiments. RJ and SKS carried out the MEG data processing and analysis. SV and SG grew and maintained the cell cultures, advised on microscopy experimental design, carried out the microscopy experiments, analyzed microscopy data, and prepared cell cultures for the MEG experiments. RJ and TMD wrote the manuscript.

**Funding**

Funding for this work was provided by NYU Abu Dhabi.

**Author Information**

To whom correspondence may be addressed:

**Ramesh Jagannathan**

**Email:** rj31@nyu.edu

**Phone:** +971 2 6284164

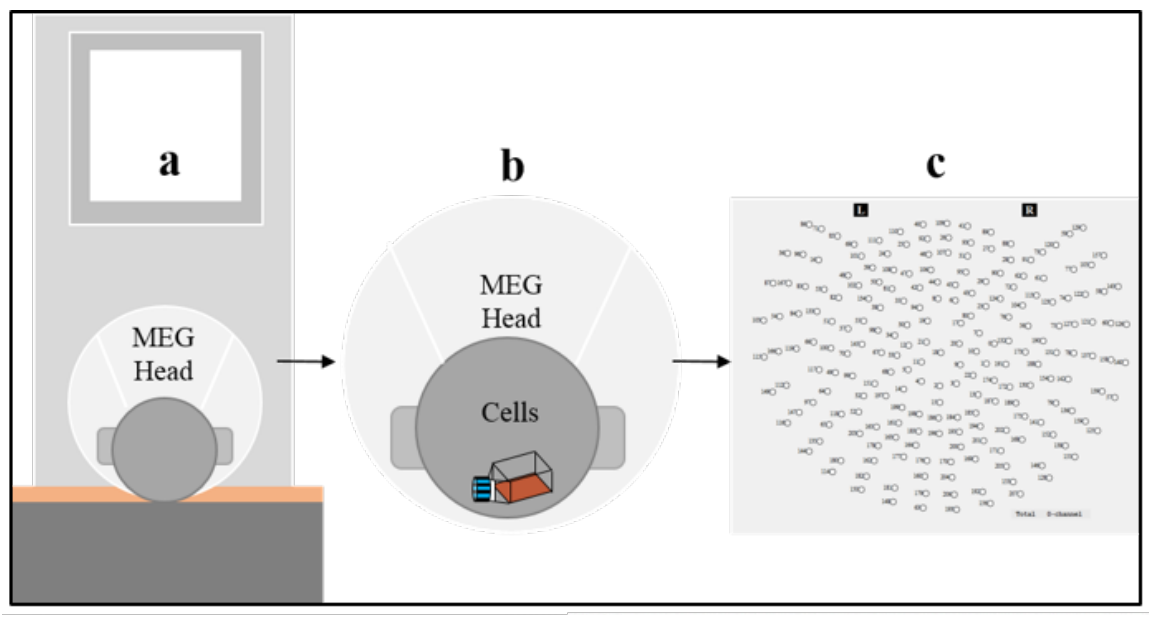

**Figure 1: Concept diagram of the MEG system used in the study and HeLa results**

(a) Sketch of the system with the MEG head, where the sensor array is located.

(b) Magnified sketch of the MEG head with the cavity where the cell culture flask is placed.

(c) Actual schematic of the sensor array located within the MEG head.

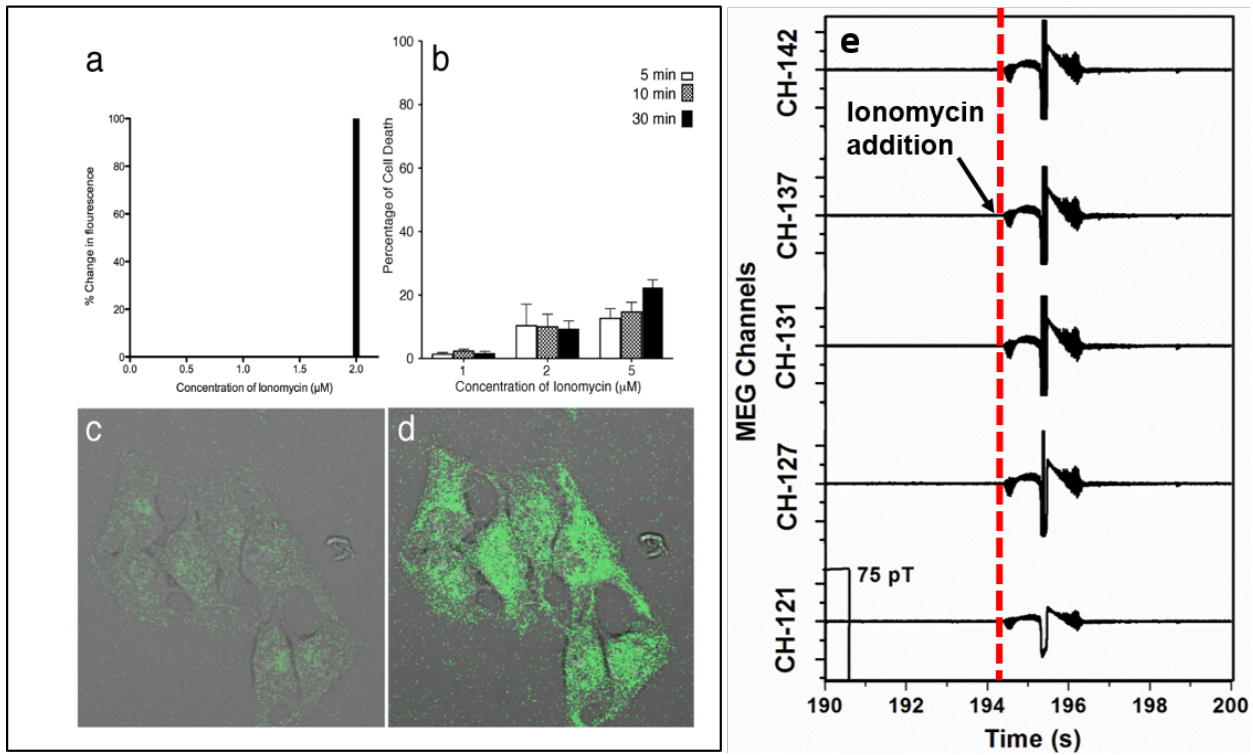

**Figure 2: Effect of ionomycin on cytosolic Ca²⁺, cell death, and the magnetic signal.** (a) Normalized fluorescent signal from Fluo-4 upon exposure to 0.5, 1.0, 1.5, and 2.0 µM ionomycin to establish the minimum ionomycin concentration to observe a response from the calcium indicator. (b) Percent of cell death from 5, 10, and 30 min exposure to 1, 2, and 5 µM ionomycin observed in a trypan blue assay. Error bars represent the standard deviation of the measurement. (c and d) Confocal images of HeLa cells loaded with Fluo-4 (c) before and (d) after addition of ionomycin (2 µM). (e) MEG data for $1 \times 10^6$ cells detected by channels 121, 127, 131, 137, and 142 when ionomycin (2 µM) is added to the flask as indicated by the dashed line.

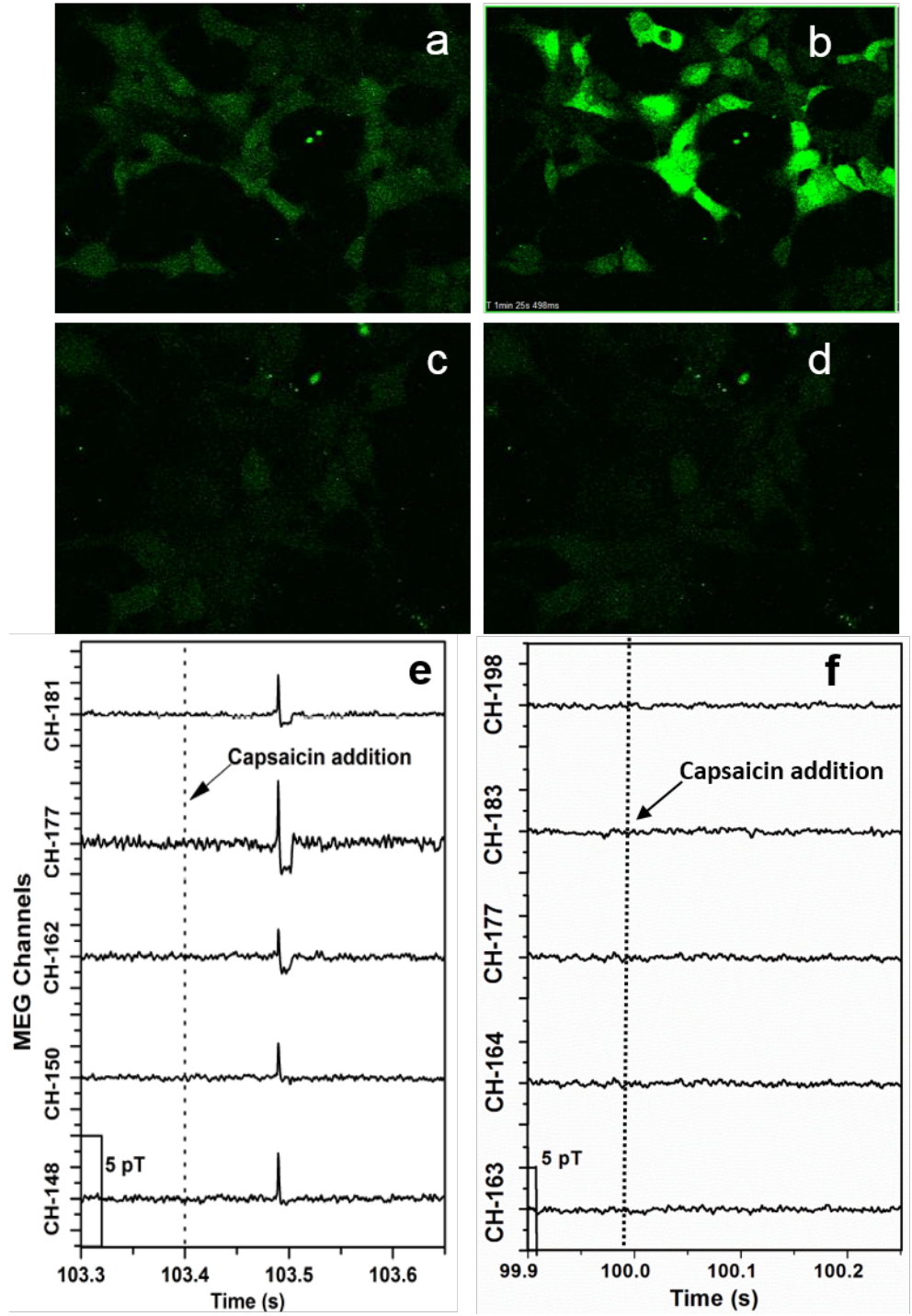

**Figure 3: Effect of capsaicin addition on cytosolic Ca²⁺ on HEK293 cells stably transfected with TRPV1 with and without the TRPV1 antagonist ruthenium red using confocal florescence imaging and MEG:** (a) Images of HEK293 cells stably transfected with TRPV1

channels loaded with Fluo-4 before addition of capsaicin. (b) Images of HEK293 cells stably transfected with TRPV1 channels loaded with Fluo-4 after addition of capsaicin (10 μM). (c) Images of HEK293 cells stably transfected with TRPV1 channels, containing the TRPV1 antagonist ruthenium red, (10 μM) loaded with Fluo-4 before addition of capsaicin (10 μM). (d) Images of HEK293 cells stably transfected with TRPV1 channels, containing TRPV1 agonist (ruthenium red, 10 μM) loaded with Fluo-4 after addition of capsaicin (10 μM). (e) MEG data for $0.05 \times 10^6$ HEK293 cells stably transfected with TRPV1 channels detected by channels 148, 150, 162, 177, and 181 when capsaicin in culture medium (10 μM final concentration) was added to the flask at the time point marked by the dotted line. (f) MEG data for $0.05 \times 10^6$ HEK293 cells stably transfected with TRPV1 channels, containing TRPV1 agonist (ruthenium red, 10μm), detected by channels 163, 164, 177, 183, and 198 when capsaicin in culture medium (10 μM final concentration) was added to the flask at the time point marked by the dotted line.

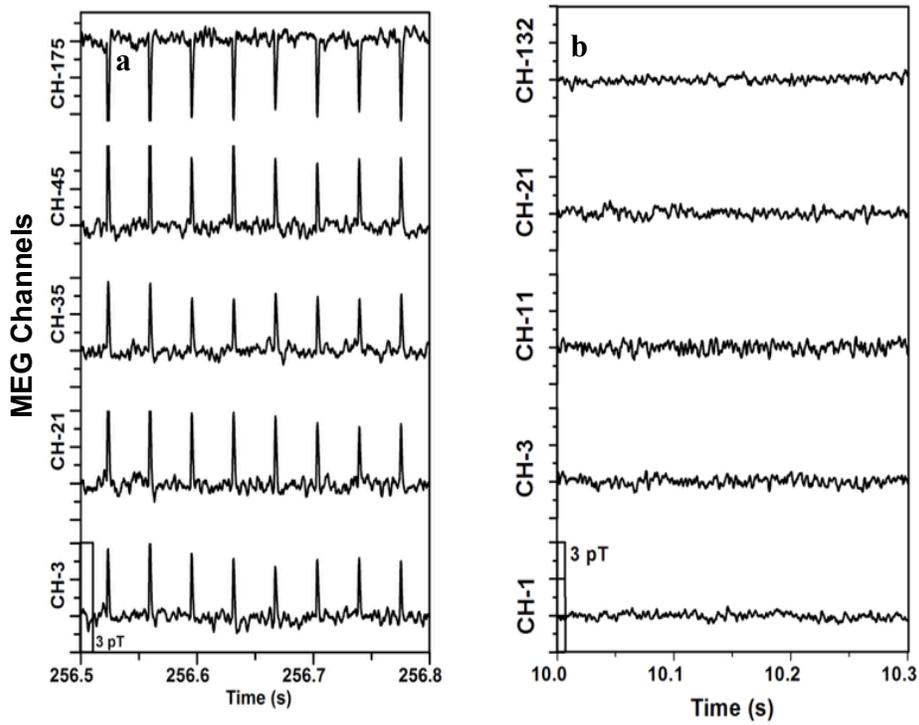

**Figure 4:** *MEG detection of magnetic fields from cells in culture* - **HeLa cells.**
(a) Typical data for $1 \times 10^6$ cells detected by channels 3, 21, 35, 45, and 175. (b) MEG data from a flask containing only culture, but no cells.

**Figure Legends**

**Figure 1: Concept diagram of the MEG system used in the study and HeLa results** (a) Sketch of the system with the MEG head, where the sensor array is located. (b) Magnified sketch of the MEG head with the cavity where the cell culture flask is placed. (c) Actual schematic of the sensor array located within the MEG head.

**Figure 2: Effect of ionomycin on cytosolic Ca$^{2+}$, cell death and on the magnetic signal.** (a) Normalized fluorescent signal from Fluo-4 upon exposure to 0.5, 1.0, 1.5, and 2.0 µM ionomycin. (b) Percent of cell death from 5, 10, and 30 min exposure to 1, 2, and 5 µM ionomycin. Error bars represent the standard deviation of the measurement. (c and d) Confocal images of HeLa cells loaded with Fluo-4 (c) before and (d) after addition of ionomycin (2 µM). (e) MEG data for $1 \times 10^6$ cells detected by channels 121, 127, 131, 137, and 142 when ionomycin (2 µM) is added to the flask as indicated by the dashed line.

**Figure 3: Effect of capsaicin addition on cytosolic Ca$^{2+}$ on HEK293 cells stably transfected with TRPV1 with and without the presence TRPV1 agonist (Ruthenium Red) using Confocal florescence imaging and MEG:** (a) Images of HEK293 cells stably transfected with TRPV1 channels loaded with Fluo-4 before addition of capsaicin. (b) Images of HEK293 cells stably transfected with TRPV1 channels loaded with Fluo-4 after addition of capsaicin (10µm). (c) Images of HEK293 cells stably transfected with TRPV1 channels, containing TRPV1 agonist (Ruthenium Red, 10µm) loaded with Fluo-4 before addition of capsaicin (10µm). (d) Images of HEK293 cells stably transfected with TRPV1 channels, containing TRPV1 agonist (Ruthenium Red, 10µm) loaded with Fluo-4 after addition of capsaicin (10µm). (e) MEG data for $0.05 \times 10^6$ HEK293 cells stably transfected with TRPV1 channels detected by channels 148, 150, 162, 177, and 181 when capsaicin in culture medium (10 µM final concentration) was added to the flask at the time point marked by the dotted line. (f) MEG data for $0.05 \times 10^6$ HEK293 cells stably transfected with TRPV1 channels, containing TRPV1 agonist (Ruthenium Red, 10µm), detected by channels 163, 164, 177, 183, and 198 when capsaicin in culture medium (10 µM final concentration) was added to the flask at the time point marked by the dotted line.

**Figure 4:** *MEG detection of magnetic fields from cells in culture -* **HeLa cells.**

(a) Typical data for $1 \times 10^6$ cells detected by channels 3, 21, 35, 45, and 175.

(b) MEG data from a flask containing only culture, but no cells.

**For Table of Contents Only**

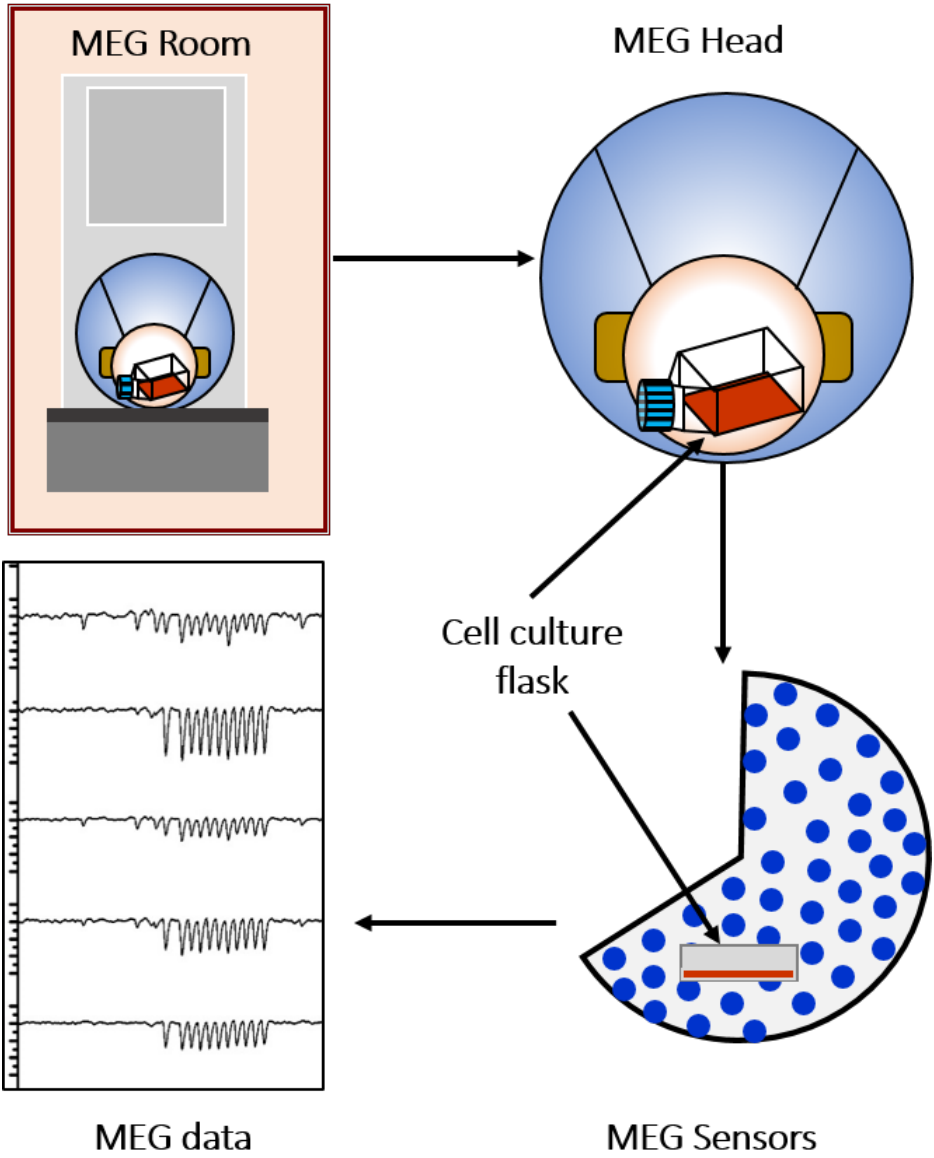

# Measuring Cellular Ion Transport by Magnetoencephalography (MEG)

Sudhir Kumar Sharma,[a] Sauparnika Vijay,[b] Sangram Gore,[b] Timothy M. Dore,[b,c]* and Ramesh Jagannathan[a]*

[a] Engineering Division, New York University Abu Dhabi, PO Box 129188, Abu Dhabi, UAE
[b] Science Division, New York University Abu Dhabi, PO Box 129188, Abu Dhabi, UAE
[c] Department of Chemistry, University of Georgia, Athens, Georgia 30602, USA

**ORCID ID**

Sudhir Kumar Sharma: 0000-0003-2437-5934

Timothy M. Dore: 0000-0002-3876-5012

Ramesh Jagannathan: 0000-0003-0269-6446

# Supplementary Figures

# Figure S1a:

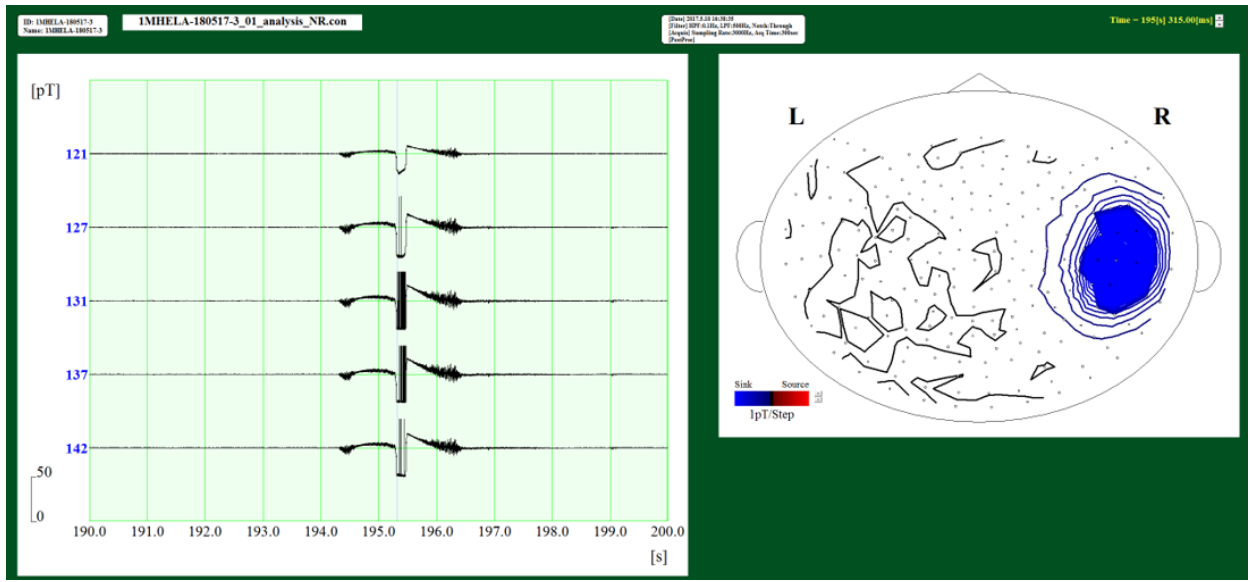

**Figure S1a:** Screen shot of noise reduced data shown in Figure 2e.

**Figure S1b:**

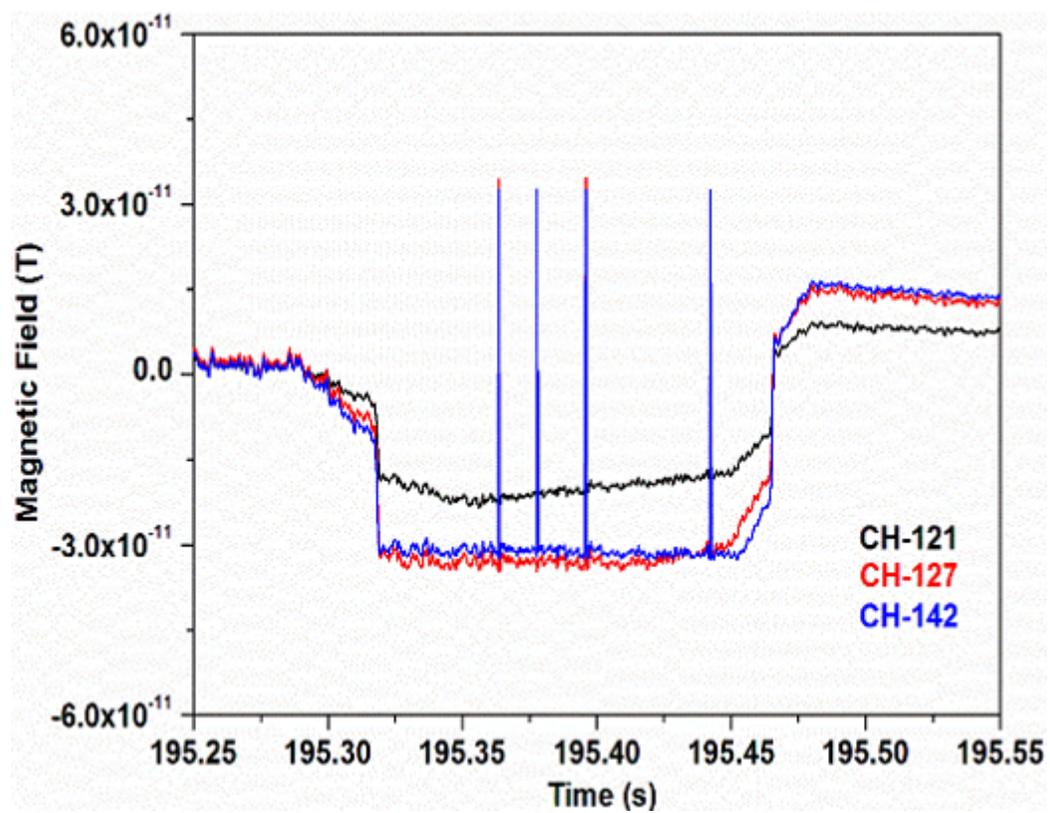

**Figure S1b:** Overlay plots of data shown in Figure 2e.

**Figure S1c:**

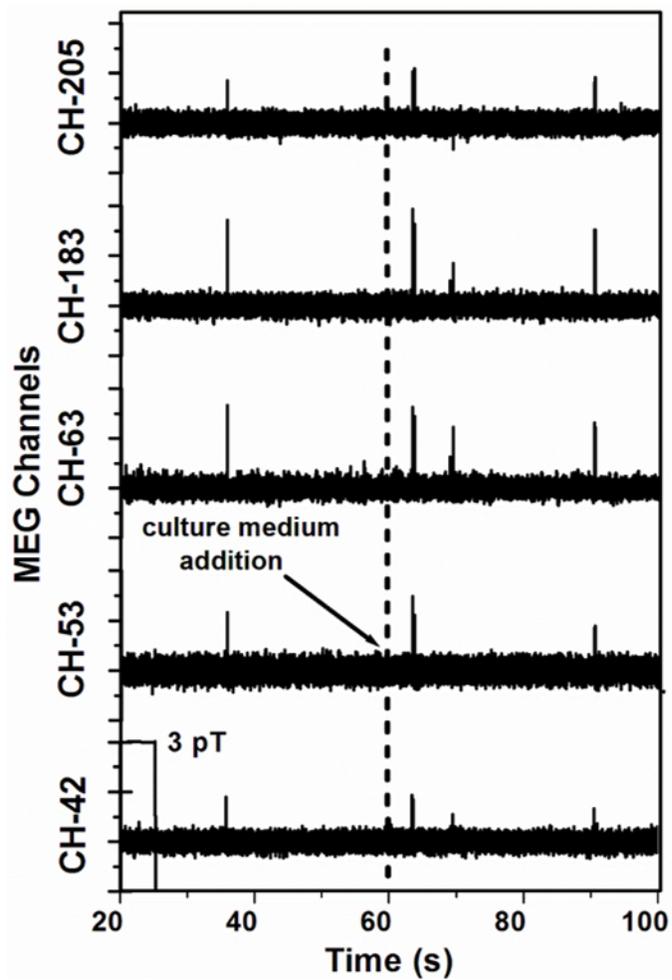

**Figure S1c:** Control experiment for the data shown in Fig. 2e. Culture medium added remotely to the cell flask.

**Figure S2a:**

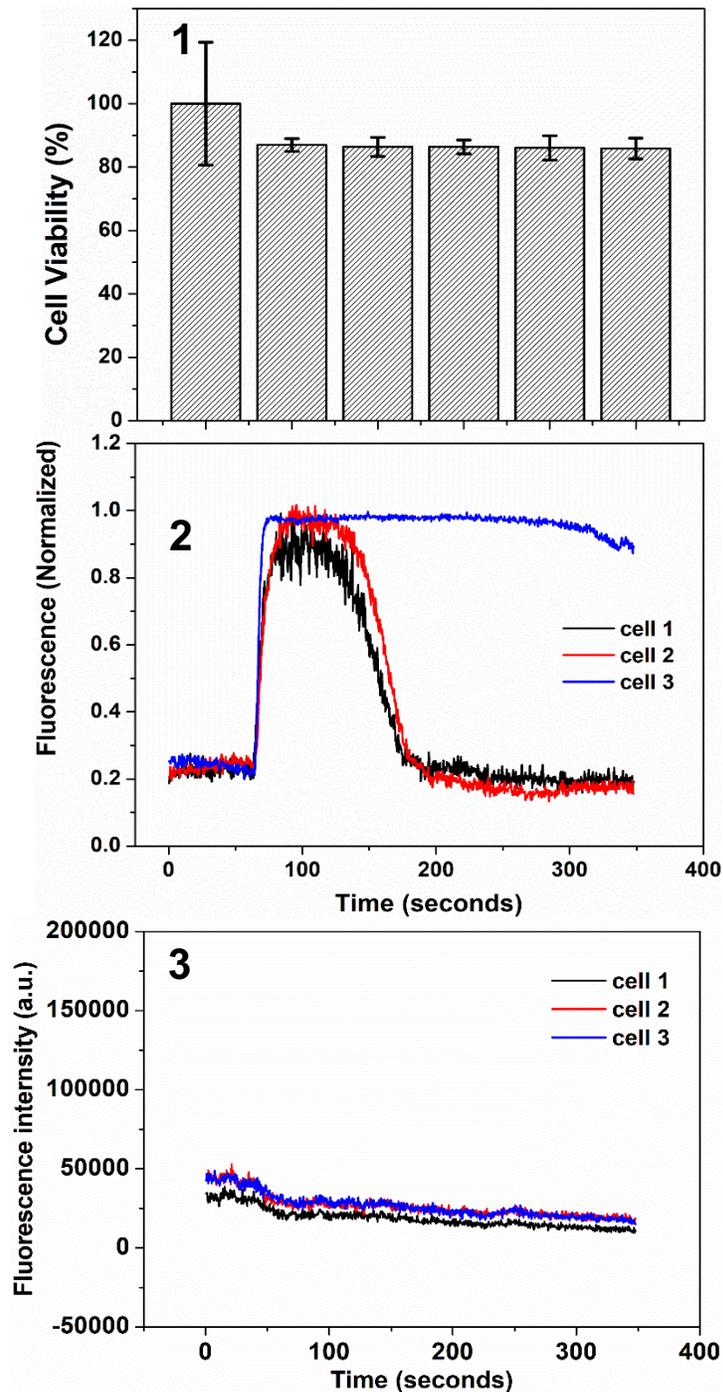

**Figure S2a:** (1) Cell viability results as a function of dosage level of capsaicin (2) Effect of addition of capsaicin (10 μM final concentration) at time (t=30 sec) to HEK293 cells expressing TRPV1 channels. (3) Effect of addition of capsaicin (10 μM final concentration) at time (t=24 sec) to HEK293 cells expressing TRPV1 channels pre-added TRPV1 antagonist (10 μM ruthenium red).

**Figure S2b:**

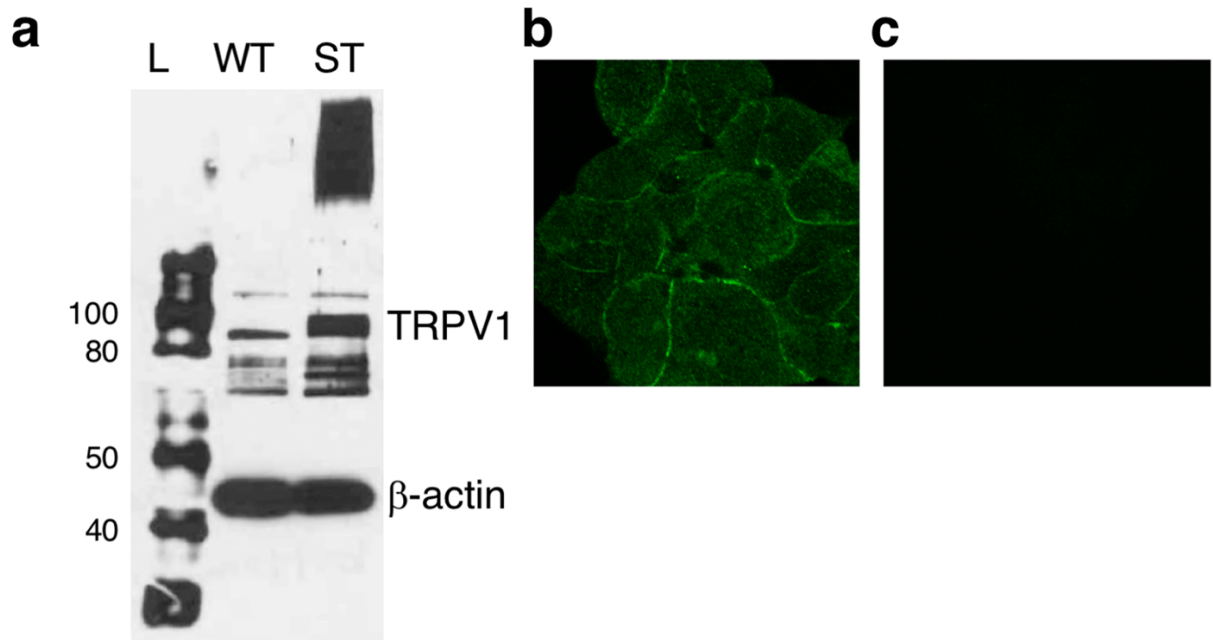

**Figure S2b:** Verification of TRPV1 (MW = 93-95 kDa) expression in stably transfected HEK 293 cells. (a) Western blot analysis of wild type (WT) and stably transfected (ST) cells and a molecular weight ladder (L): protein extracts were loaded on 10% SDS polyacrylamide gels, then the nitrocellulose membranes were split at the position of proteins of 40 kDa. The antibodies used and their dilutions were anti-TRPV1 (1:500, rabbit polyclonal, Origene, cat. #TA309921), anti-biotin (1:1,000, HRP linked), and anti-rabbit-HRP (1:2,000, goat secondary). β-Actin protein (MW = 42 kDa) was used as a control. (b and c) Immunofluorescence staining of HEK 293(TRPV1) cells treated (b) with TRPV1 specific antibody (anti-TRPV1, 1:250; secondary antibody, 1:500) and (c) without primary antibody.

**Figure S2c:**

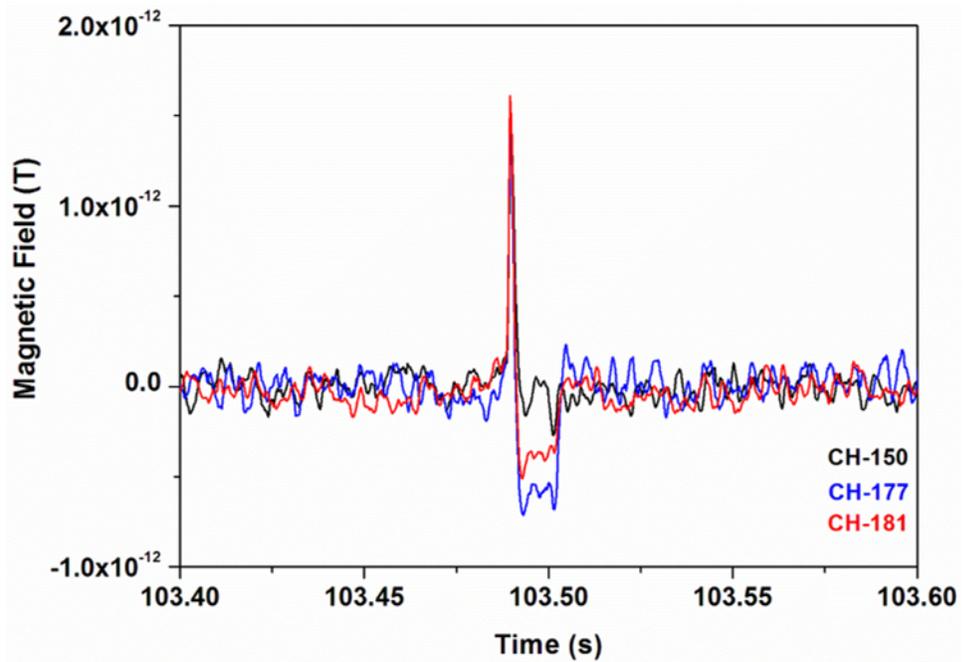

**Figure S2c:** Overlay plots of data shown in Fig, 3e.

**Figure S2d:**

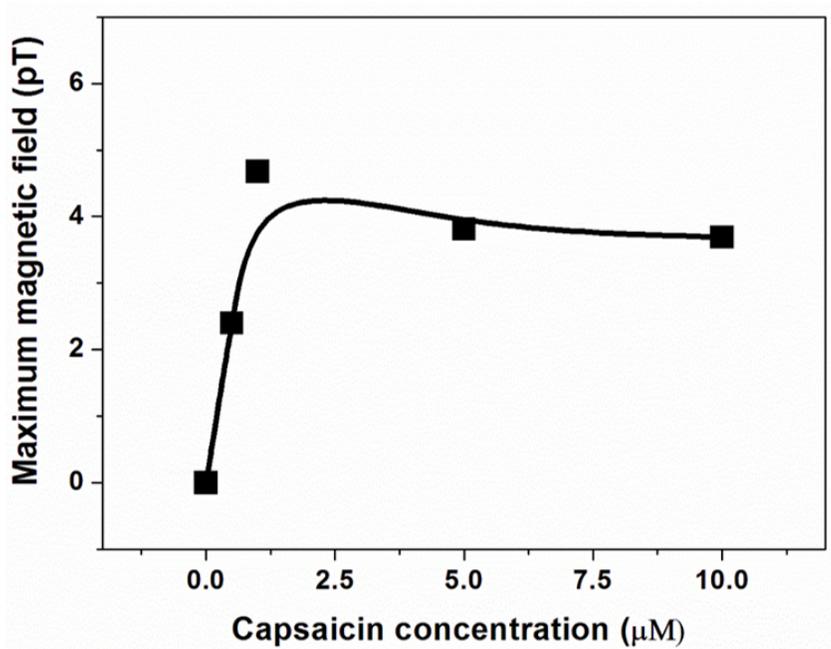

**Figure S2d:** Effect of increasing [capsaicin] concentration on the measured magnetic field from TRPV1-expressing HEK293 cells.

# Figure S2e:

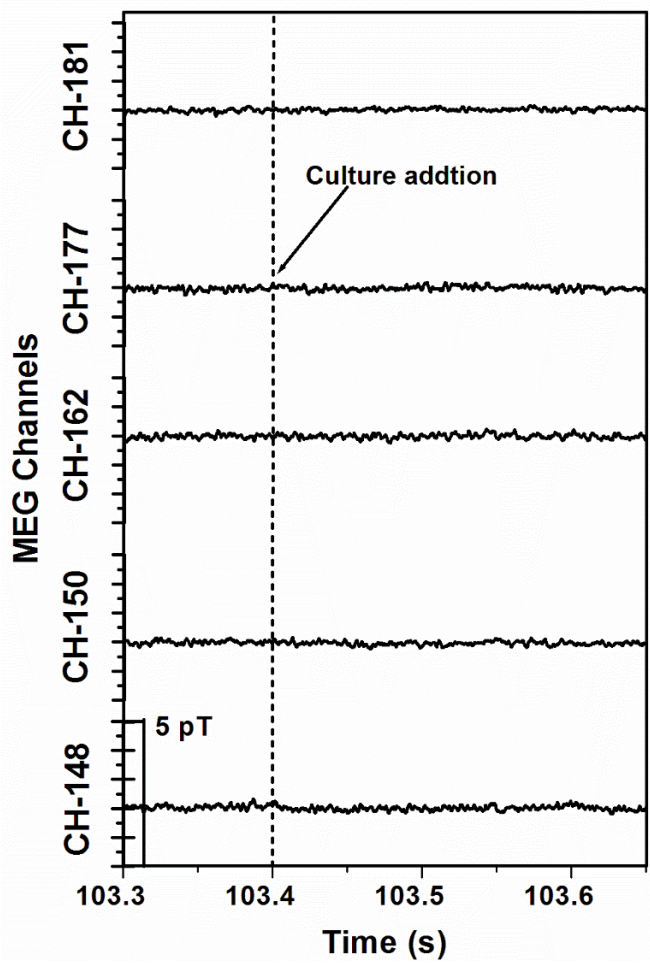

**Figure S2e:** Control experiment for the data shown in Fig. 3e. Effect of addition of culture without capsaicin remotely to the cell flask.

**Figure S2f:**

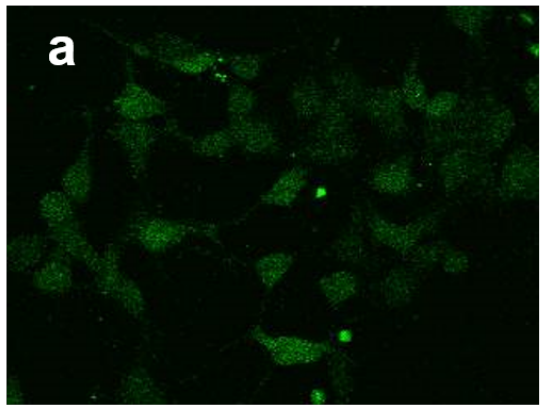

Before Capsaicin Addition

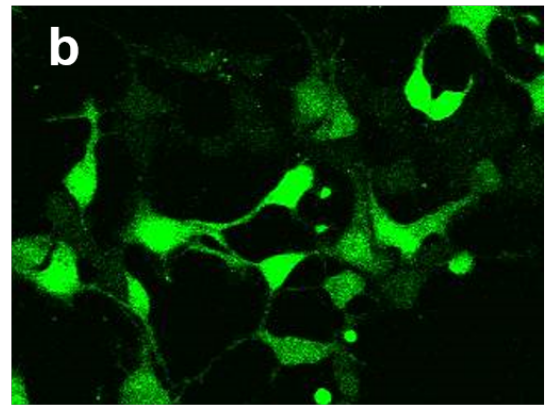

After 1st Capsaicin Addition

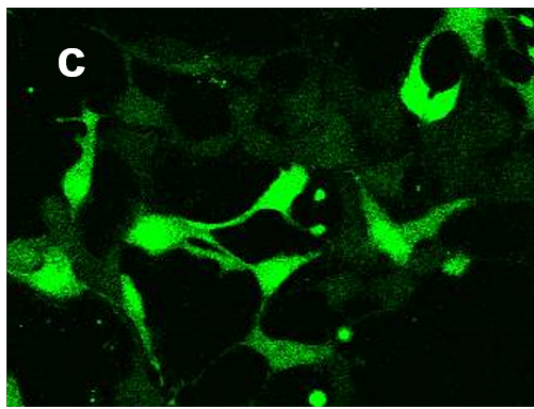

After 2nd Capsaicin Addition

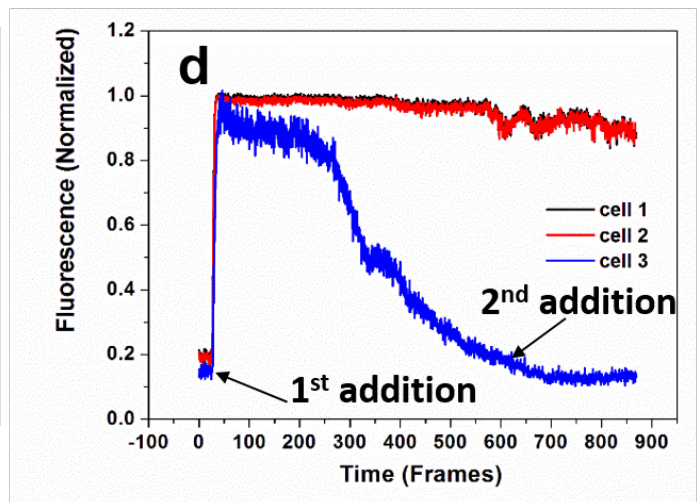

**Figure S2f:** Confocal fluorescence experimental data on the effect of multiple additions of capsaicin to the same TRPV1-expressing HEK293 cells: (a) before capsaicin addition, (b) after capsaicin addition, (c) after second capsaicin addition, (d) fluorescence (normalized) as a function of time, after the first capsaicin addition and after the second capsaicin addition.

**Figure S2g:**

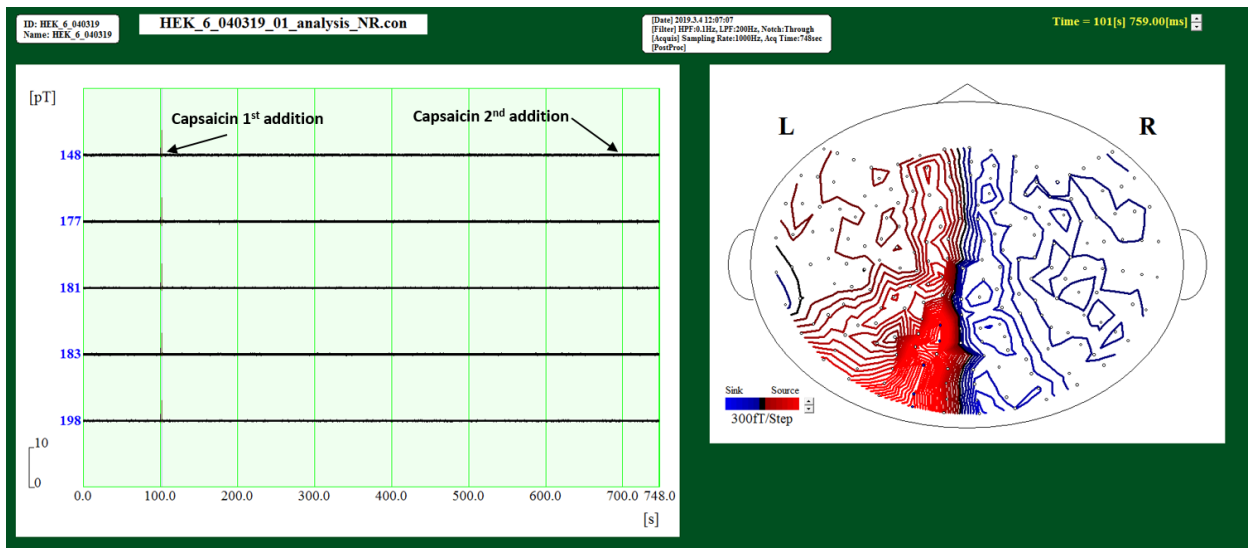

**Figure S2g:** MEG noise reduced data on the effect of multiple additions of Capsaicin to the same HEK293 cells expressing TRPV1channels.

**Figure S2h:**

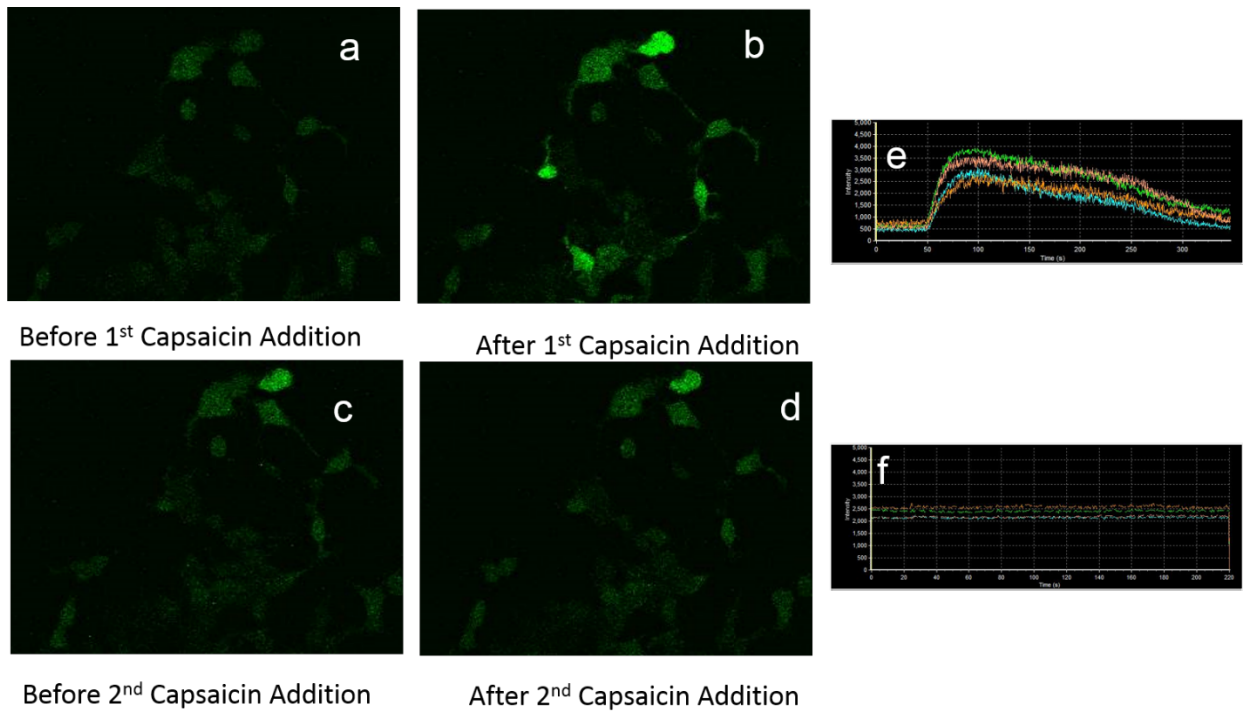

Before 1st Capsaicin Addition

After 1st Capsaicin Addition

Before 2nd Capsaicin Addition

After 2nd Capsaicin Addition

**Figure S2h:** Confocal fluorescence experimental data on the effect of replacing the culture before the second capsaicin addition, by washing the with fresh culture three times: (a) before first capsaicin addition, (b) after first capsaicin addition, (c) before second capsaicin addition, (d) after second capsaicin addition, (e) fluorescence intensity vs time after first addition (screen shot), (f) fluorescence vs time after second addition (screen shot).

**Figure S2i:**

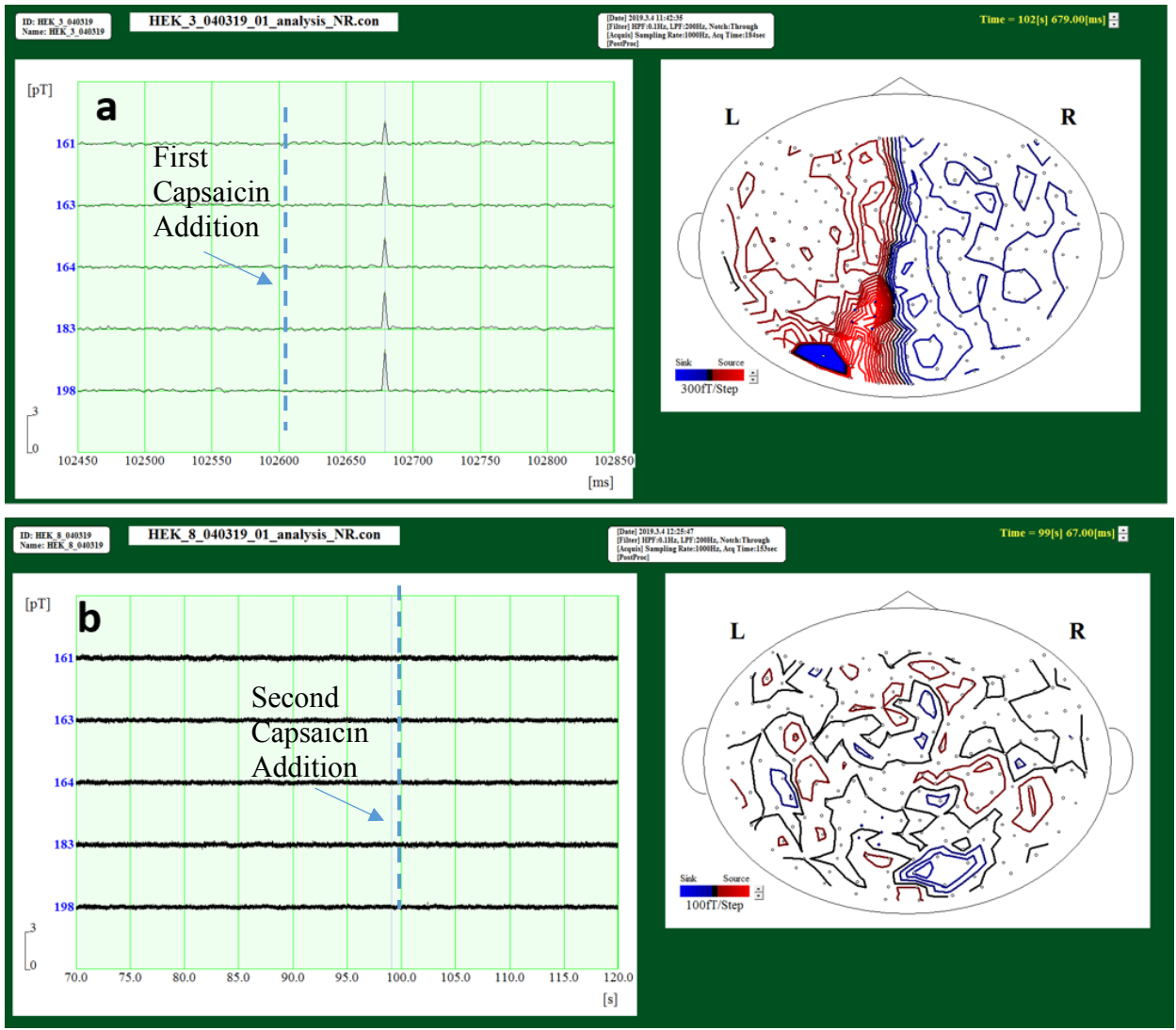

**Figure S2i:** MEG data on the effect of replacing the culture before the second capsaicin addition, by washing the TRPV1-expressing HEK293 cells with fresh culture three times. a) Screen shot of the noise reduced data for first capsaicin addition. b) Screen shot of noise reduced data for the second capsaicin addition.

# Figure S3a:

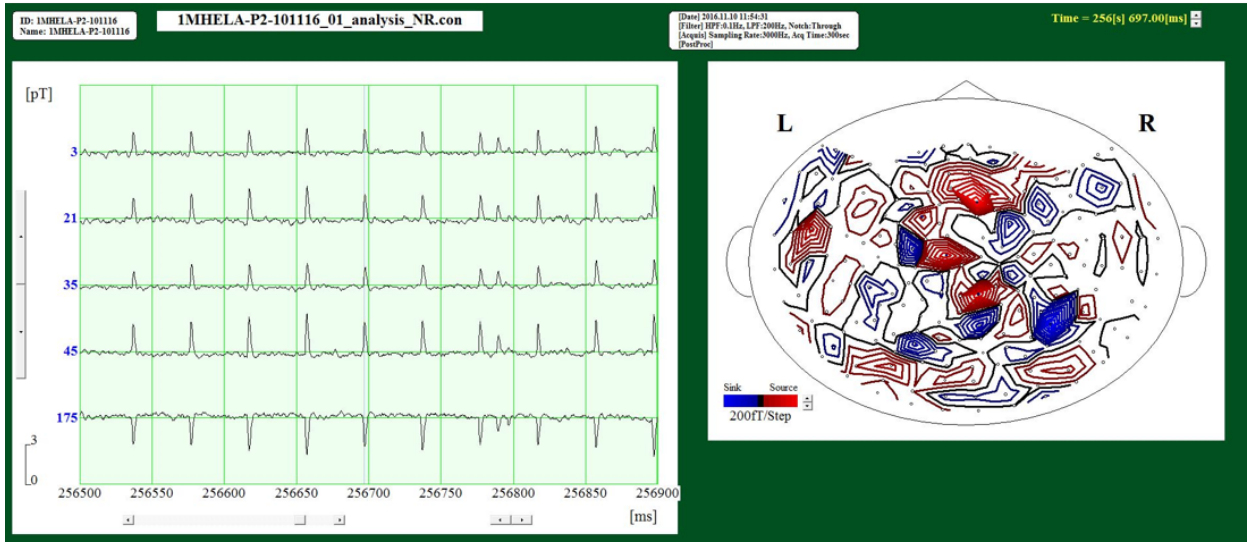

**Figure S3a:** Screen shot of the MEG noise reduced data shown in Fig. 4a.

**Figure S3b:**

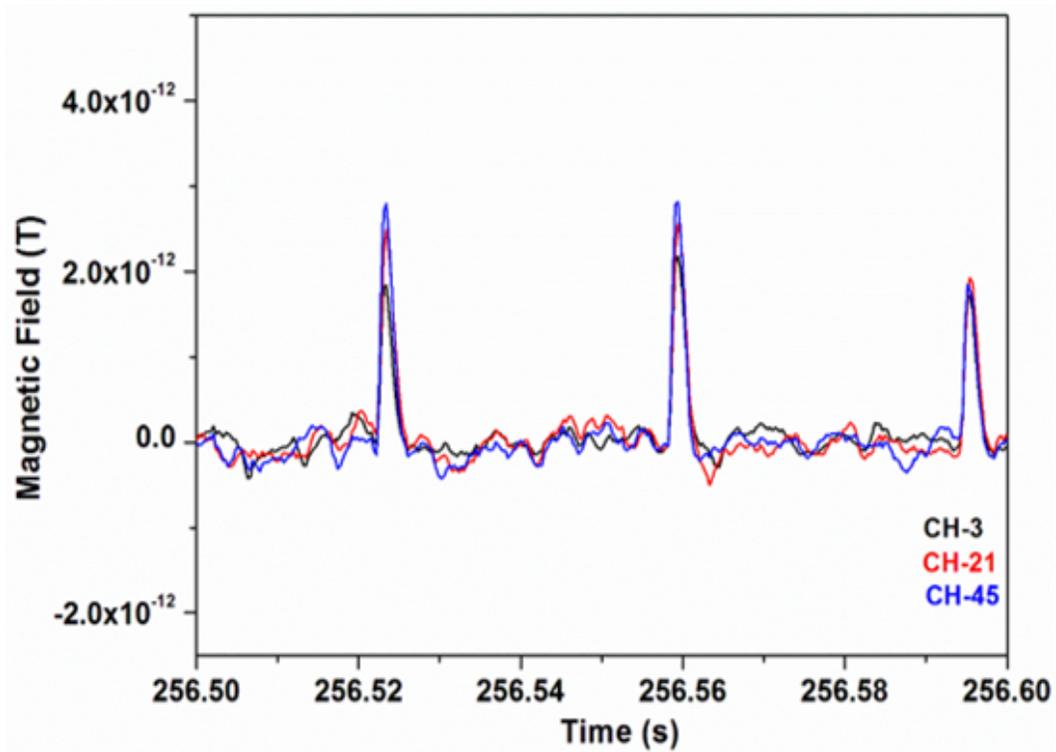

**Figure S3b:** Overlay plots of data shown in Fig. 4a.

**Figure S3c:**

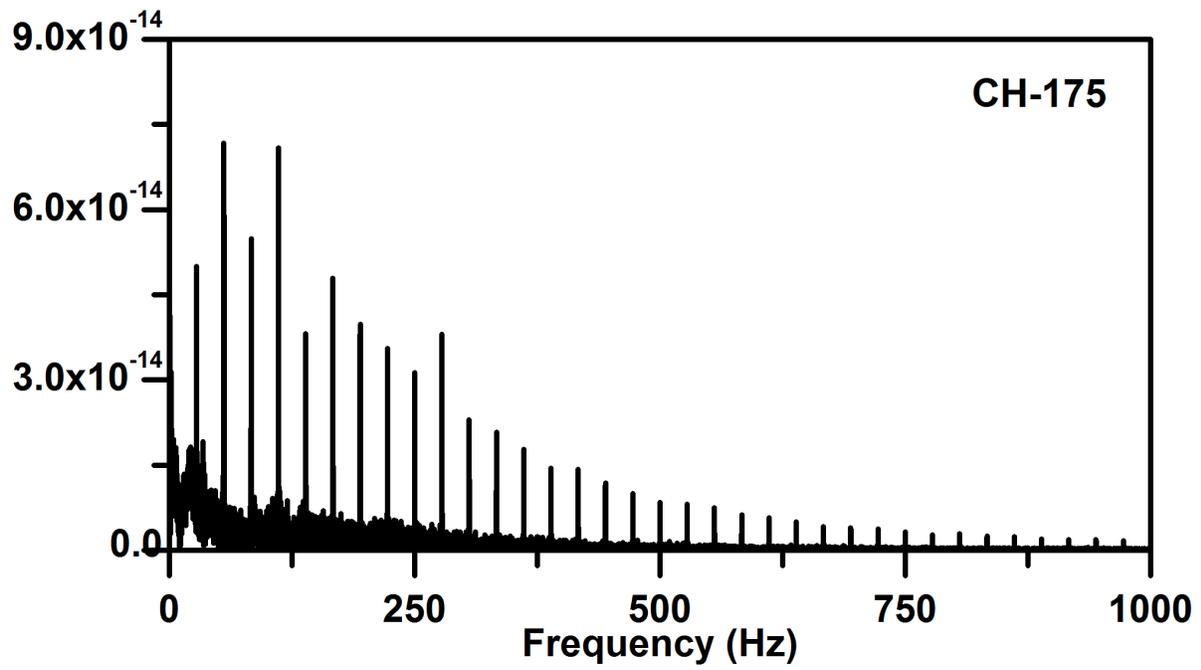

**Figure S3c:** FFT of the data of channel 175 shown in Figure 4.

**Figure S3d:**

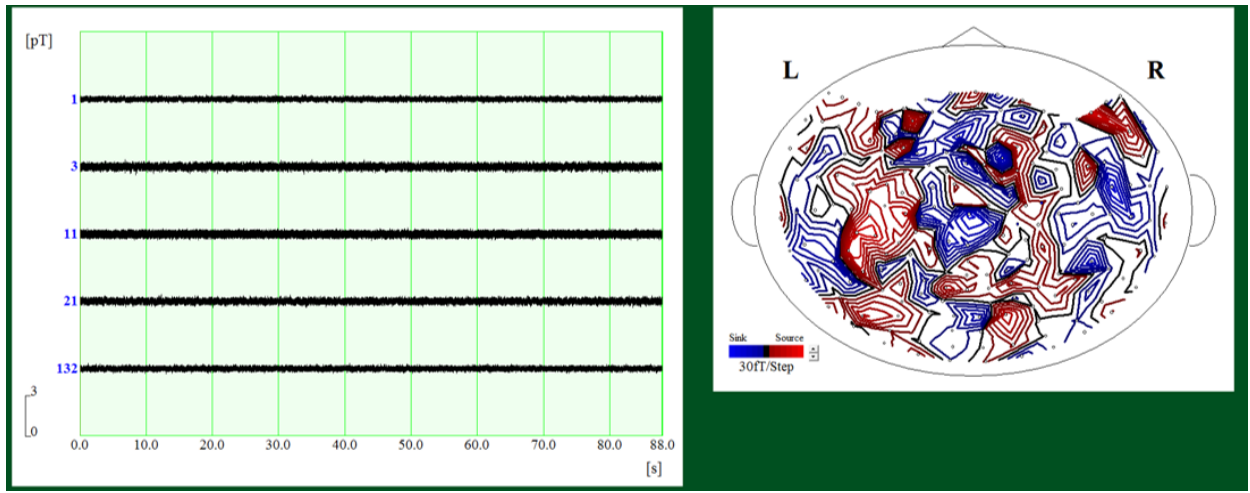

**Figure S3d:** Screen shot of the MEG noise reduced signals from the control experiment data shown in Fig.4b.

**Figure S3e:**

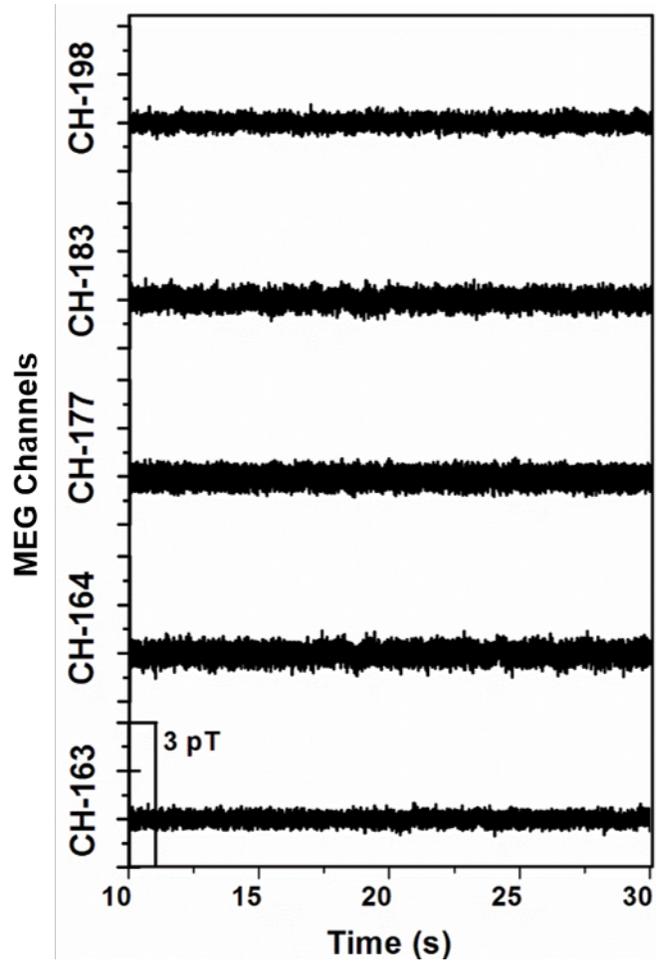

**Figure S3e:** Magnetic fields from HeLa cells dispersed in culture.

# Figure S4:

**a**

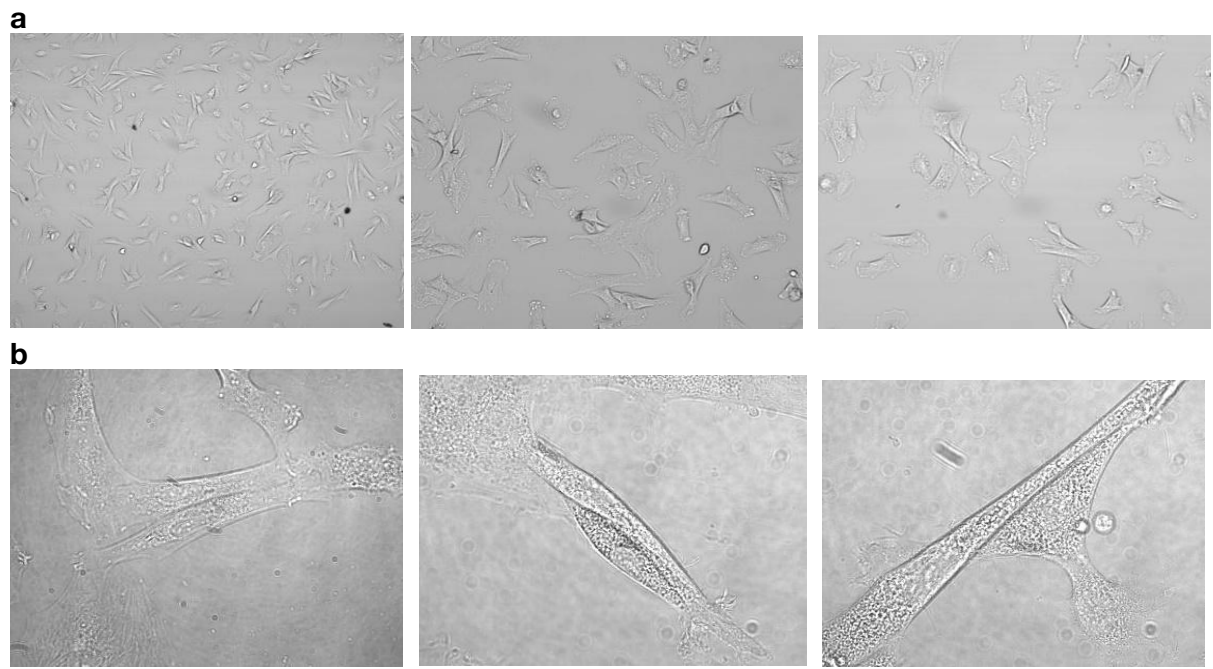

**b**

**Figure S4: Morphological characterization of H9c2(2-1) cells before and after differentiation.** (a) Representative images of non-differentiated H9c2(2-1) cells grown in 10% FBS media showing the characteristic stellar shape of myoblasts. (b) Representative images of H9c2(2-1) cells after five days of differentiation in low serum media (2% FBS) showing the characteristic elongated shape of myocytes.

# Figure S5:

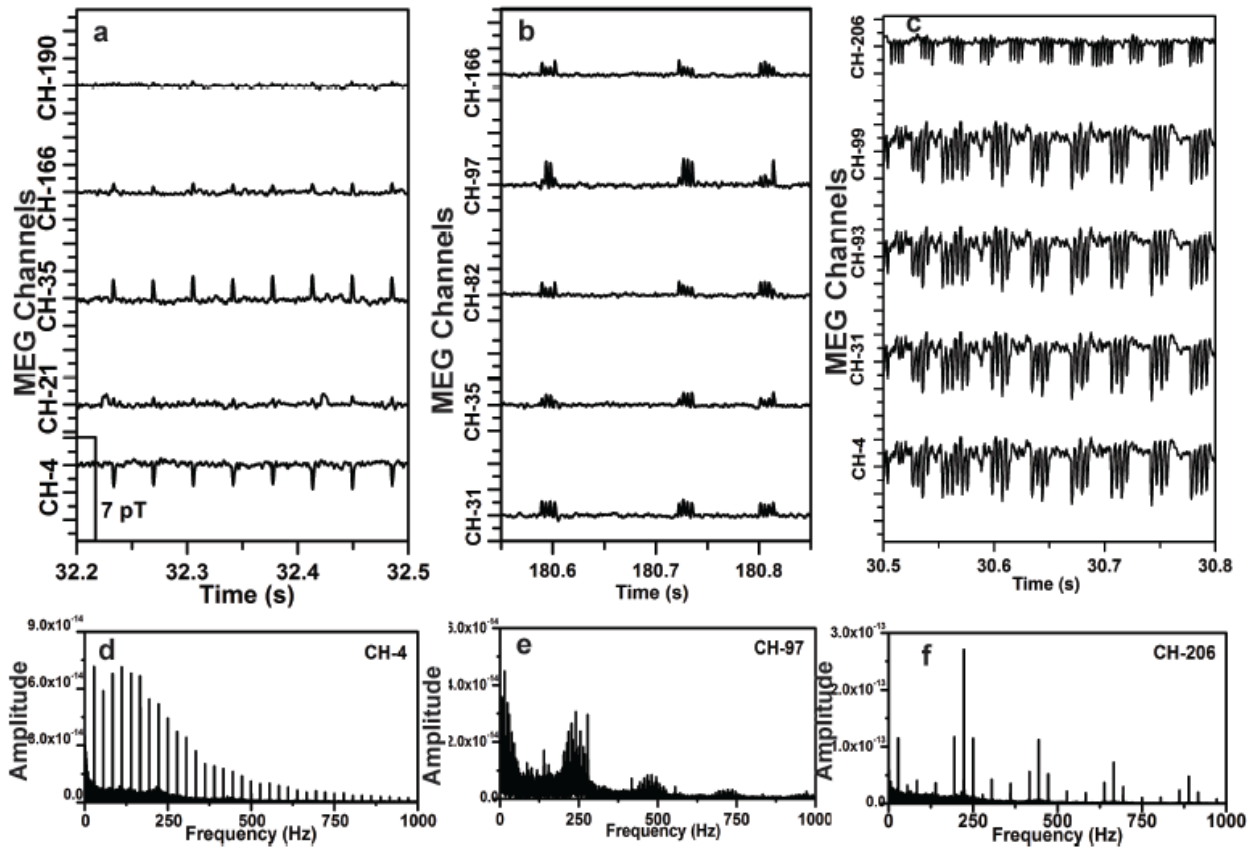

**Figure S5: Comparison of the magnetic signals from non-differentiated and differentiated cardiac cells.** (a) Data for $1 \times 10^6$ H9c2(2-1) rat cardiac myoblasts detected by channels 4, 21, 35, 166, and 190. (b and c) Data for $1 \times 10^6$ H9c2(2-1) rat cardiac cells differentiated into myocytes detected by channels 31, 35, 82, 97, and 166 and 4, 31, 93, 99, and 206. (d) FFT of the data for non-differentiated cells (channel 4). (e and f) FFT of the data for differentiated cells (channels 97 and 206).